\documentclass[preprint,aps,amsmath,superscriptaddress,nofootinbib,tightenlines]{revtex4}
\usepackage{graphicx}
\usepackage{bm}
\usepackage{epsfig}

\def\bsigma{\mbox{\boldmath $\sigma$}}

\def\OMIT#1{}

\newcommand{\nn}{\nonumber}

\newcommand{\bn}{{\bar n}}
\newcommand{\bea}{\begin{eqnarray}}
\newcommand{\eea}{\end{eqnarray}}
\newcommand{\nb}{\bar n}

\newcommand{\bnP}{\bar {\cal P}}
\newcommand{\ppP}{{\cal P}_\perp}

\newcommand{\cP}{{\cal P}}

\newcommand{\mpsi}{M_\psi}

\newcommand{\jpsi}{J/\psi}
\newcommand{\emax}{E_{max}}

\begin{document}

%



\title{The color-singlet contribution to $e^+ e^- \to \jpsi + X$ at the endpoint}
	
\author{Adam K. Leibovich\footnote{Electronic address: akl2@pitt.edu}}
\author{Xiaohui Liu\footnote{Electronic address: xil41@pitt.edu}}
\affiliation{Department of Physics and Astronomy,
	University of Pittsburgh,
        Pittsburgh, PA 15260\vspace{0.2cm}}

\date{\today\\ \vspace{1cm} }



\begin{abstract}
Recent observations of the $J/\psi$ spectrum produced in $e^+e^-$ collisions at the $\Upsilon(4S)$ resonance are in conflict with fixed-order calculations using the Non-Relativsitic QCD effective theory(NRQCD).  One problem is an enhancement in the cross section when the $J/\psi$ has maximal energy, due to large perturbative corrections (Sudakov logarithms).  In a recent paper, the Sudakov logarithms in the color-octet contribution were summed by combining NRQCD with the Soft-Collinear Effective Theory.  However to be consistent, the color-singlet contributions must also be summed in the endpoint region which was not done in that paper.  In this paper, we sum the leading and next-to-leading logarithms in the color-singlet contribution to the $J/\psi$ production cross section.  We find that the color-singlet cross section is suppressed near endpoint compared to the fixed order NRQCD prediction.
\end{abstract}

\maketitle

\newpage

\section{Introduction}

Bound states of heavy quarks and antiquarks have been of great interest
since the discovery of the $\jpsi$~\cite{psi}. In particular
the production of quarkonium is an interesting probe of both perturbative
and nonperturbative aspects of QCD dynamics.
Production  requires the creation of a heavy  $Q\bar{Q}$ pair with energy
greater than $2 m_Q$, a scale at which  the
strong coupling constant is small enough that perturbation theory can be used.
However, hadronization
probes much smaller mass scales of order $m_Q v^2$, where $v$ is the typical velocity
of the quarks in the quarkonium. For $J/\psi$, $m_Q v^2$ is numerically of order $\Lambda_{\rm QCD}$ so the
production  process is sensitive to nonperturbative physics as well.

Many phenomenological problems can be understood well enough by using the Non-Relativistic Quantum Chromodynamics (NRQCD)~\cite{Bodwin:1995jh, Luke:2000kz}. NRQCD provides a generalized factorization theorem that includes nonperturbative corrections to the color-singlet model. All infrared divergences can be factored into nonperturbative matrix elements, so that infrared safe calculations of inclusive decay rates are possible~\cite{Bodwin:1992ye}.  However, there are some predictions of NRQCD are in conflict with the data, in particular the predicted polarization of $J/\psi$ at the Fermilab Tevatron~\cite{psipol, poldata} and more recently the production rate of $J/\psi$ associated with extra $c$ and $\bar c$ quarks (both inclusive and exclusive) at the $B$ factories \cite{Abe:2002rb,Aubert:2005tj}.  In particular, Belle reports a
large cross section for $J/\psi$ produced along with open charm~\cite{Abe:2002rb},
\begin{eqnarray}
\frac{\sigma (e^+e^- \rightarrow J/\psi c \bar{c})}{\sigma (e^+e^- \rightarrow J/\psi X)} =0.59^{+0.15}_{-0.13}\pm 0.12 \nonumber \, .
\end{eqnarray}
The predicted ratio from leading order color-singlet production mechanisms alone is about $0.2$ \cite{Cho:1996cg,Baek:1998yf} 
 and a large color-octet contribution makes this ratio even smaller. 
In addition to the inclusive measurements, Belle reports a cross section for exclusive double charmonium 
production which exceeds previous theoretical estimates. Recent attempts to address the latter problem can be found in Ref.~\cite{double}.

The inclusive $J/\psi$ production at the $B$ factories is another potential conflict between experimental observations and theoretical predictions using NRQCD~\cite{Abe:2001za, Aubert:2001pd}.
 Leading order NRQCD calculations predict that for most of the range of allowed energies prompt $J/\psi$ production should be dominated by color-singlet production mechanisms, while color-octet contributions dominate when the $J/\psi$ energy is nearly maximal. Furthermore, as pointed out in  Ref.~\cite{Braaten:1995ez}, color-octet processes predict a dramatically different  angular distribution for the $J/\psi$. Writing the differential cross section as
\begin{eqnarray}
\frac{d \sigma}{d p_\psi \, d \cos\theta} = S(p_\psi)[1+ A(p_\psi) \cos^2 \theta] \, ,
\end{eqnarray}
where $p_\psi$ is the $J/\psi$ momentum and $\theta$ is the angle of the $J/\psi$ with respect to the axis defined by the $e^+e^-$ beams, one finds the color-singlet mechanism gives $A(p_\psi)\approx 0$ except for large $p_\psi$, where $A(p_\psi)$ becomes large and negative. On the other hand, color-octet production predicts $A(p_\psi) \approx 1$. The significant enhancement of the  cross section accompanied by the change in angular distribution were proposed as a distinctive signal of color-octet mechanisms in  Ref.~\cite{Braaten:1995ez}. It was expected that these effects would be confined to $J/\psi$ whose momentum is within a few hundred  MeV of the maximum allowed.

However, experimental results do not agree with these expectations. The cross section data as a function of momentum does not exhibit any enhancement in the bins closest to the endpoint. On the other hand, the total cross section measured by the two experiments exceeds predictions based on the color-singlet model alone. The total prompt $J/\psi$ cross section, which includes feeddown from $\psi^\prime$ and $\chi_c$ states but not from $B$ decays, is measured to be $\sigma_{tot} = 2.52 \pm 0.21 \pm 0.21$ pb by BaBar, while Belle measures $\sigma_{tot} = 1.47 \pm 0.10 \pm 0.13$ pb. Estimates of the color-singlet contribution range from $0.4 -0.9$ pb \cite{Cho:1996cg,Yuan:1996ep,Baek:1998yf,Schuler:1998az}. Furthermore,  $A(p_\psi)$ is measured to be consistent with $1$ (with large errors) for $p_\psi > 2.6 \,{\rm GeV}$ (Belle)  and $p_\psi > 3.5 \,{\rm GeV}$ (BaBar).

The NRQCD factorization formalism shows that the differential $\jpsi$ cross section can be written as
\begin{equation} \label{NRQCDprod}
d \sigma (e^+ e^- \rightarrow \jpsi + X) = \sum_n d
\hat{\sigma} (e^+ e^- \rightarrow c \bar{c}[n]+ X) \langle {\cal O}^{\jpsi}_n \rangle \,,
\end{equation}
where $d \hat{\sigma}$ is the inclusive cross section for producing a
$c\bar{c}$ pair in a color and angular momentum state labeled by $[n] = {}^{2S+1}L_J^{(i)}$. In this notation,
the spectroscopic notation for angular momentum quantum numbers is standard and
$i = 1 (8)$ for color-singlet (octet) production matrix elements.
The short-distance coefficients are
calculable in a perturbation series in $\alpha_s$. The
long-distance matrix elements $\langle {\cal O}^{\jpsi}_n \rangle$ are
vacuum matrix elements of four-fermion operators in
NRQCD~\cite{Bodwin:1995jh}. These matrix elements scale as some power
of the relative velocity $v \ll 1$ of the $c$ and $\bar{c}$ quarks as
given by the NRQCD power-counting rules.

At lowest order in $v$ the only term in Eq.~(\ref{NRQCDprod}) is the color-singlet
contribution, $[n]={}^3S_1^{(1)}$, which scales as $v^3$.  The coefficient for this
contribution starts at $O(\alpha^2_s)$ \cite{csrate}.  There are two different contributions to the leading-order color-singlet, depending on what else is produced along with the $J/\psi$: $e^+e^-\to J/\psi+g+g$ and $e^+e^-\to J/\psi +c+\bar c$. 
Away from the kinematic endpoint $\emax = (s+\mpsi^2)/ (2 \sqrt{s})$,
where $s$ is the center-of-mass energy squared, color-octet contributions also start
at $O(\alpha_s^2)$. Since the color-octet contributions are suppressed by $v^4 \sim 0.1$
relative to the leading color-singlet contributions, they are negligible
throughout most of the allowed phase-space at leading order in perturbation theory.  

The theoretical situation becomes more interesting, however, near the endpoint.  The lowest-order, color-singlet term approaches a constant\footnote{The $J/\psi c\bar c$ contribution goes to zero before the kinematic endpoint, due to the non-zero mass of the charm quarks.  It therefore does not contribute to the endpoint contribution given in Eq.~(\ref{csendpoint}).}
\begin{equation}\label{csendpoint}
\lim_{z\rightarrow 1}\frac{d \hat{\sigma}[{}^3S_1^{(1)}]}{d z \, d \cos{\theta}} =
\frac{64 \pi \alpha^2 \alpha^2_s e^2_c}{27 s^2 m_c}(1+r)
\Bigg( \frac{1+r}{1-r} - \cos^2{\theta} \Bigg) \,.
\end{equation}
where $r = 4 m^2_c /s$, and $z=E_{c\bar{c}}/E^{max}_{c\bar{c}}$ with $E^{max}_{c\bar{c}} = \sqrt{s}(1+r)/2$, while the lowest-order, color-octet piece is singular (proportional to a delta function).  Physically, when the $J/\psi$ emerges with close to the maximal energy, it is recoiling against an energetic gluon jet with energy of order $M_\Upsilon$ but invariant mass of order $M_\Upsilon \sqrt{ \Lambda_{\rm QCD}/M_\psi}$.  The degrees of freedom needed to describe this inclusive jet have been integrated out of NRQCD, and thus cannot be described by the effective field theory.  The effective theory which correctly describes this kinematic regime is a combination of NRQCD for the heavy degrees of freedom, and the Soft-Collinear Effective Theory (SCET)~\cite{Bauer:2001ew,Bauer:2001yr,Bauer:2001ct,Bauer:2001yt} for the light energetic degrees of freedom.  Furthermore, the renormalization group equations of SCET will sum the large kinematic preturbative corrections which appear near corners of phase space.

In a previous paper~\cite{Fleming:2003gt} the combination of NRQCD and SCET was used to sum the large kinematic logarithms (Sudakov logarithms) which appear in the color-octet contribution near the endpoint.  For the color-octet contribution, there are also large non-perturbative contributions at the endpoint~\cite{Beneke:1997qw} which must also be summed into a non-perturbative shape function.  Since the shape function is unknown, in Ref.~\cite{Fleming:2003gt} the shape function was modeled.  Since it is universal, it is possible that it could be extracted from another process (such as $J/\psi$ photoproduction~\cite{Fleming:2006cd}).  With the summation of the perturbative corrections and the simple model chosen, a good fit to the data was obtained.  

However, to be consistent, the color-singlet contribution should also be summed in the endpoint region.  This is the goal of the present paper.  The kinematic logarithms in the $J/\psi +c+\bar c$ color-singlet contribution are small, since the mass of the charm quark acts as a cutoff.  However, we would expect that the summed $J/\psi +g+g$ color-singlet rate would be suppressed relative to the unsuppressed rate.  This would help alleviate the discrepancy with the open charm data.  However, we would not expect a very large suppression except right near the endpoint, and thus do not expect that this will be a solution to the $J/\psi\ +$ open charm question.  This will be confirmed in our analysis in this paper. 
The remainder of the paper is organized as follows.  In Sec II a factorization theorem for $J/\psi$ production near the endpoint is developed.  Then in Sec III the Sudakov logarithms are summed, including mixing with the $\jpsi+q+\bar q$ final state.  In Sec IV the phenomenology of the $J/\psi$ production is investigated, and finally we conclude in Sec V.
A similar treatment of nonperturbative and perturbative endpoint corrections to the color-singlet and color-octet contributions in the inclusive decay $\Upsilon \rightarrow X +\gamma$ can be found in Refs.~\cite{Fleming:2002sr,Fleming:2004rk}, and we will rely on some of the results from these papers.  Similar results have been previously reported in Ref.~\cite{Lin:2004eu}.


\section{Factorization}\label{fact}

In this section, we will derive a factorization theorem for $e^+ e^- \to \jpsi + X$ near the kinematic endpoint, where the rate can be factored into a hard coefficient, a collinear jet function and a ultrasoft shape function.  The derivation is quite similar to Refs.~\cite{Bauer:2001yt,Fleming:2002sr,Fleming:2003gt,Fleming:2004rk,Fleming:2006cd,Bauer:2002nz}.
We begin by briefly reviewing the kinematics of the process in the $e^+ e^-$ center of mass (COM) frame\cite{Fleming:2003gt}. In the COM frame, the virtual photon has momentum $q^\mu = \sqrt{s}/2(n^\mu+\bn^\mu)$ with the lightlike vectors defined as $\bn^\mu = (1,0,0,1)$ and $n^\mu=(1,0,0,-1)$. The $\jpsi$ is moving in the $z$-direction with four-velocity
\begin{equation}
v^{\mu}=\frac{1}{2} \Bigg( \frac{M_{\psi}}{x \sqrt{s}} n^\mu +  \frac{x \sqrt{s}}{M_{\psi}} \bn^\mu  \Bigg) \,.
\end{equation}
Here $M_{\psi}$ is the $\jpsi$ mass and  $x  = (E_{\psi}+p_{\psi})/\sqrt{s}$. The $c\bar{c}$ pair has momentum $p_{c\bar{c}}^{\mu}= Mv^{\mu}+\ell^{\nu} = Mv^{\mu}+\Lambda^{\mu}{}_{\nu} \hat{\ell}^{\nu}$, where $M = 2 m_c$ and $\ell^{\nu}$ is the residual momentum of the $c\bar{c}$ pair inside the $J/\psi$.  In the $J/\psi$ rest frame, $\hat\ell^\mu$ has components of $O(\Lambda_{\rm QCD})$, which get boosted in the COM frame to $\ell^\mu$ scaling as $ \nb \cdot \ell \sim M_\psi\Lambda_{\rm QCD}/(x\sqrt{s}), n \cdot \ell \sim x \sqrt{s}\Lambda_{\rm QCD}/M_\psi$ and $  \ell_\perp   \sim  \Lambda_{\rm QCD}$.  The momentum of the gluon jets is
\begin{equation}\label{jetmom}
p^\mu_X = \frac{\sqrt{s}}{2} \Bigg[ \Bigg(1 - \frac{r}{\hat{x} } \Bigg) n^\mu  +(1-\hat{x}) \bar{n}^\mu \Bigg]  - \ell^\mu \, ,
\end{equation}
where  $\hat{x} = x M /M_\psi$.  In the end point region the NRQCD factorization formula breaks down because NRQCD does not include appropriate collinear modes.  When $1-x \sim  \Lambda_{\rm QCD}/M$, the jet is no longer highly virtual. Since $m_X^2/E_X^2 \sim \Lambda_{\rm QCD}/M \ll 1$, the gluon jet is composed of energetic particles with small invariant mass which must be included explicitly in the effective theory. Hence, a new factorization theorem is needed to handle the end point,  which can be derived using a combination of NRQCD for the heavy quark degrees of freedom and  SCET\cite{Bauer:2001ct, Bauer:2001ew, Bauer:2001yr, Bauer:2001yt} which includes the collinear physics.

SCET has collinear degrees of freedom whose momentum scales as  $\nb \cdot p \sim Q$,  $n \cdot p \sim \lambda^2 Q$, and $p^\perp\sim \lambda Q$, soft degrees of freedom whose momentum scales as $\lambda$ and ultrasoft (usoft) degrees of freedom whose momentum scales as $\lambda^2$. Heavy quark fields in SCET are the same as in NRQCD when considering quarkonium.  For $e^+e^-\to\jpsi+X$,   $Q$ is of order $\sqrt{s}$, while $\lambda \sim \sqrt{1-x}\sim \sqrt{\Lambda_{\rm QCD}/M}$. To the order we are working, operators will contain usoft, collinear quarks and gluons and heavy quark fields. Soft fields do not enter to the order we are interested and are neglected.

We match QCD onto SCET at the scale $Q$ by evaluating matrix elements in QCD at the scale $Q$ and expanding in powers of $\lambda$.  Each order in $\lambda$ is reproduced in the effective theory by the product of SCET operators and Wilson coefficients.  All the dependence on the large scale $Q$ shows up in the Wilson coefficients.  We must  include all SCET operators which can contribute to the process under consideration at each order of $\lambda$. These operators must respect the symmetries of the effective theory. For $e^+e^- \rightarrow J/\psi + X$, the operators must be invariant under both collinear and usoft gauge transformations \cite{Bauer:2001yt}. Lorentz invariance is realized in the effective theory by additional constraints on the operators, called reparametrization invariance (RPI) \cite{Manohar:2002fd}.

In the collinear sector of SCET there is a collinear fermion field $\xi_{n,p}$, a collinear gluon field $A_{n,q}^\mu$(soft modes are ignored), and a collinear Wilson line
\begin{equation}
W_n(x)=
 \bigg[ \sum_{\rm perms} {\rm exp}
  \left( -g_s \frac{1}{\bnP} \bn \cdot A_{n,q}(x) \right) \bigg] \,.
\end{equation}
The subscripts on the collinear fields are the lightcone direction $n^\mu$, and the large components of the lightcone momentum ($\bn\cdot
q, q_\perp$). The operator ${\cal P}^\mu$ projects out the momentum label~\cite{Bauer:2001ct}, $\bn\cdot{\cal P}\xi_{n,p} \equiv \bnP\xi_{n,p} = \bn\cdot p\xi_{n,p}$.  In the usoft sector there is a usoft fermion field $q_{us}$, a usoft gluon field $A^\mu_{us}$, and a usoft Wilson line $Y$. Using the transformation properties for each of these
fields under collinear and usoft gauge transformations~\cite{Bauer:2001yt}, we can build invariant operators.
The collinear-gauge invariant field strength is
\begin{equation}\label{colfieldstrength}
G^{\mu\nu}_n \equiv -\frac{i}{g_s} W^\dagger [i{\cal D}_n^\mu
  + g_sA_{n,q}^\mu, i{\cal D}_n^\nu+g_sA_{n,q'}^\nu ] W ,
\end{equation}
where
\begin{equation}\label{cov}
i{\cal D}_n^\mu = \frac{n^\mu}2 \bnP + \ppP^\mu +
\frac{\bn^\mu}2 i n\cdot D,
\end{equation}
and $iD^\mu = i \partial^\mu+g_sA^\mu_{us}$ is the usoft covariant
derivative.  RPI requires the label operators and the usoft
covariant derivatives, which scale  differently with $\lambda$,
to appear in the linear combination appearing in $i{\cal D}_n^\mu$.
The leading piece of $G^{\nu \mu}_n$ is order $\lambda$ and
can be written as $\bn_\nu G^{\nu\mu}_n = i[\bnP ,  B^\mu_\perp]$, where
\begin{equation}\label{bfield}
B^\mu_\perp =  \frac{1}{g_s} W^\dagger (\ppP^\mu + g_s (A^\mu_{n,q})_\perp)W.
\end{equation}
The subscript $\perp$ on $B_\perp^\mu$ indicates that $\mu$ must be a perpendicular direction.

\begin{figure}[t]
\centerline{ \includegraphics[width=4in]{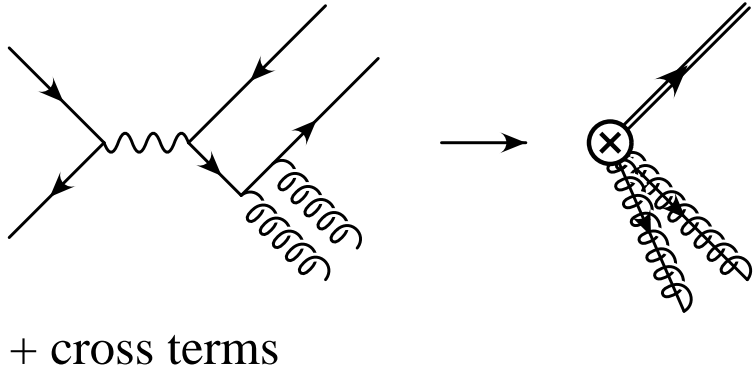}}
\caption{\it Matching the production amplitude for $e^+ e^- \rightarrow c\bar{c}+gg$ in QCD and SCET. Collinear gluons are represented by a spring with a line through it.
\label{prod}}
\end{figure}

We next construct the operators necessary to describe color-singlet ${}^3S_1$ production at the end point. A $c\bar{c}$ pair in a color-singlet ${}^3S_1$ configuration must be accompanied by a colorless jet of quarks and gluons. The leading operator must have two gluon field operators to create the collinear gluons in the final state. Thus, we should construct the operator out of two $B_{\perp}$ fields in color singlet configuration. Taking gauge-invariance into consideration, the only operator is
\begin{equation}\label{3s1op1}
{\cal O}_{\mu\; gg}(1,{}^3S_1) =
 \chi^\dagger_{-{\bf p}} \Lambda\cdot\bsigma^\delta \psi_{\bf p}
{\rm Tr} \big\{ B^\alpha_\perp \,
\Gamma^{(1,{}^3S_1)}_{\alpha \beta \delta \mu} ( \bnP, \bnP^\dagger ) \,
B^\beta_\perp \big\} \,.
\end{equation}
At leading order, the coefficient is determined by requiring the SCET matrix element of Eq.~(\ref{3s1op1}) to reproduce the lowest order QCD diagrams for $e^+ e^- \rightarrow c\bar{c}+gg$, shown in Fig.~\ref{prod}.
Matching at tree level, we obtain
\begin{equation}\label{matching1}
\Gamma^{(1,{}^3S_1)}_{\alpha\beta\mu\delta} = \frac{32\pi}{3}\frac{e_{c}e\alpha_s}{M^2}\frac{r}{1-r}g^{\perp}_{\alpha\beta} \Big(g_{\mu\delta}-\frac{1-r}{2}n_{\mu}n_{\delta} \Big) \,,
\end{equation}
where $r = 4 m^2_c /s$ and  $g_\perp^{\mu\nu} = g^{\mu\nu} - (n^\mu\bar n^\nu + n^\nu\bar n^\mu)/2$.  We can also have a jet made up of a quark-antiquark pair.  Again, taking gauge-invariance into account, the only operator is
\begin{equation}\label{3s1op1qq}
{\cal O}_{\mu\;\bar qq}(1,{}^3S_1) =
 \chi^\dagger_{-{\bf p}} \Lambda\cdot\bsigma^\delta \psi_{\bf p}
\bar\xi_{n,p} W_n
\Gamma^{(1,{}^3S_1)}_{\delta \mu} ( \bnP, \bnP^\dagger ) \,
W_n^\dagger \xi_{n,p} \,.
\end{equation}
The leading order Wilson coefficient is zero.  However, since this operator occurs at the same order in $\lambda$, it can be generated through mixing.  Just as in the case of Ref.~\cite{Fleming:2004rk}, the mixing is small, and we will neglect this term for now.

At leading order in the SCET power counting the cross section in the endpoint can be expressed in a factored form to all orders in $\alpha_s$
\begin{equation}\label{factthmI}
2 E_\psi \frac{d\sigma}{d^3p_\psi} = \frac{e^2}{16\pi^3 s^3} L^{\mu\nu} H_{\mu\nu} \int dl^+ S(l^+, \mu) J_{\omega}(l^+-\sqrt{s}(1-\hat{x})) \,,
\end{equation}
where $J$ is the collinear jet function, $S$ is the usoft function and $H_{\mu\nu}$ is the hard coefficient.  We shall now prove this factorization theorem.  Using the optical theorem, the production cross section can be written as
\begin{eqnarray}\label{xsec}
2 E_\psi \frac{d\sigma}{d^3p_\psi} &=& \frac{e^2}{16\pi^3 s^3}
L^{\mu\nu}
\sum_X \langle 0 | J^\dagger_\nu(0) | J/\psi+X\rangle \langle J/\psi+X | J_\mu(0) | 0\rangle
(2\pi)^4 \delta^4(q-p_\psi-p_X)\nonumber\\
&=& \frac{e^2}{16\pi^3 s^3}
L^{\mu\nu}
\int d^4 y\, e^{-i q\cdot y}
\sum_X \langle 0 | J^\dagger_\nu(y) | J/\psi+X\rangle
\langle J/\psi+X | J_\mu(0) | 0\rangle \nonumber \\
&\equiv&\frac{e^2}{16\pi^3 s^3} L^{\mu\nu} {\rm Im } T_{\mu\nu},
\end{eqnarray}
where the sum includes integration over the phase space of $X$.  The lepton tensor is
\begin{equation}
L^{\mu\nu} = p_1^\mu p_2^\nu + p_1^\nu p_2^\mu - g^{\mu\nu} p_1\cdot p_2,
\end{equation}
where $p_{1,2}$ are the momenta of the electron and positron, respectively,
and
\begin{equation}\label{Ttensor}
T_{\mu\nu} =-i\int d^4y e^{-iqy} \sum_X \langle0|J^\dagger_\mu(y)|J/\Psi+X\rangle \langle J/\Psi+X|J^\dagger_\nu(0)|0\rangle \,.
\end{equation}

The first step is to match the QCD current $J_\mu$ in Eq.~(\ref{xsec}) to leading order in $\lambda$,
\begin{equation}\label{current}
J^{\mu}=\sum_\omega e^{-i(Mv-\bnP n/2)\cdot y}\Gamma^{\alpha\beta\mu\delta}\tilde{J}_{\alpha\beta\delta}(\omega)\,,
\end{equation}
where the effective current is
\begin{equation}
\tilde{J}^{\alpha\beta\delta}=\psi^\dagger_{p}(\Lambda \cdot \boldsymbol{\sigma} )^{\delta} \chi_{-p} \{{\rm Tr}[B^{\alpha}_{\perp} \delta_{\omega \cP_{-}} B^{\beta}_{\perp}] \}\,,
\end{equation}
and $\Gamma^{(1,{}^3S_1)}_{\alpha\beta\mu\delta}(\omega)$ is given in Eq.~(\ref{matching1}). Substituting Eq.~(\ref{current}) into Eq.~(\ref{Ttensor}) and using $q^{\mu}-Mv^{\mu}+\bnP n^{\mu}/2 \approx \sqrt{s}(1-\hat{x})\nb^{\mu}/2$
gives
\begin{equation}\label{csbot}
T_{\mu \nu} =
\sum_{\omega,\omega'} \Gamma^{\dagger}_{\alpha^{\prime} \beta^{\prime} \delta^{\prime}\mu } \Gamma_{\alpha\beta\delta\nu}
T^{\alpha\alpha'\beta\beta'\delta\delta'}_{\rm eff}(\omega,\omega', \hat{x},\mu) \,,
\end{equation}
where
\begin{equation}\label{Teff}
T_{\alpha\alpha'\beta\beta'\delta\delta'}^{\rm eff} = -i\int d^4y e^{-i\sqrt{s}(1-\hat{x})\nb \cdot y} \sum_X \langle0|\tilde{J}^{\dagger}_{\alpha^{\prime}\beta^{\prime} \delta'}(\omega^{\prime})|\jpsi+X\rangle \langle \jpsi+X|\tilde{J}_{\alpha\beta\delta}(\omega)|0\rangle  \,.
\end{equation}
%
%
%

Next we decouple the usoft gluons in $T_{\rm eff}$ using the field redefinition \cite{Bauer:2001yt}
\begin{equation}\label{fieldredef}
A^\mu_{n,q} = Y A^{(0) \mu}_{n,q} Y^\dagger
\hspace{.5cm} \to \hspace{.5cm}
W_n = Y W_n^{(0)} Y^\dagger \,,
\end{equation}
where the first identity implies the second.  The collinear fields with the superscript $(0)$ do not interact with usoft fields
to lowest order in $\lambda$. In the color-singlet contribution all usoft Wilson lines $Y$ cancel due to the identity $Y^\dagger Y = 1$. Furthermore, the $\jpsi$ does not contain any collinear quanta, so using
\begin{eqnarray}\label{ucomplete}
\sum_{X_u} | J/\psi+X_u \rangle \langle J/\psi+X_u | &=&
a_\psi^\dagger \sum_{X_u} | X_u \rangle \langle X_u | a_\psi = a_\psi^\dagger a_\psi, \\
\label{ccomplete}
\sum_{X_c} | X_c \rangle \langle X_c | &=& 1,
\end{eqnarray}
where $a_\psi^\dagger a_\psi$ projects onto final states containing a $\jpsi$, we can write
\begin{eqnarray}\label{tba}
T^{\alpha\alpha'\beta\beta'\delta\delta'}_{\rm eff} &=&
\int d^4 y\, e^{-i\sqrt{s}/2(1-\hat{x})\bn\cdot y}
\langle 0 | \chi^\dagger_{-{\bf p}} (\Lambda \cdot \boldsymbol{\sigma} )^{\delta'} \psi_{\bf p} (y)
\,a_\psi^\dagger a_\psi\,
\psi^\dagger_{\bf{p}} (\Lambda \cdot \boldsymbol{\sigma} )^{\delta} \chi_{-\bf{p}}(0) | 0\rangle\\
&&\phantom{\frac12 \int d^4 y\, e^{i\sqrt{s}/2(1-x)\bn\cdot y}}
\times \langle0|\{{\rm Tr}[B_{\perp}^{\alpha^{\prime}}\delta_{\omega^{\prime} P_{-}} B_{\perp}^{\beta^{\prime}}](y) \} \{{\rm Tr}[B_{\perp}^{\alpha}\delta_{\omega P_{-}} B_{\perp}^{\beta}](0) \} |0\rangle. \nonumber
\end{eqnarray}

We can use spin symmetry to simplify the usoft matrix element,
\begin{eqnarray}
&&
\Lambda^{\delta'}_i\Lambda^\delta_j \langle 0 | \chi^\dagger_{-{\bf p}} \boldsymbol{\sigma}^i \psi_{\bf p} (y)
\,a_\psi^\dagger a_\psi\,
\psi^\dagger_{\bf{p}} \boldsymbol{\sigma}^j \chi_{-\bf{p}}(0) | 0\rangle = \nn\\
&&\qquad \frac{1}{3} \delta^{ij} \Lambda^\delta_i \Lambda^{\delta'}_j
 \langle 0 | \chi^\dagger_{-{\bf p}} \boldsymbol{\sigma}^k \psi_{\bf p} (y)
\,a_\psi^\dagger a_\psi\,
\psi^\dagger_{\bf{p}} \boldsymbol{\sigma}^k \chi_{-\bf{p}}(0) | 0\rangle.
\end{eqnarray}
Then we can use the identity $\delta^{ij} \Lambda^\delta_i \Lambda^{\delta'}_j = (v^\delta v^{\delta'} - g^{\delta \delta'})$, where $v^\delta$ is the four-velocity of the $J/\psi$, to further simplify the result.

We can define a collinear jet function from the collinear matrix element,
\begin{eqnarray}\label{jetfunction}
\langle0|\{{\rm Tr}[B^{\alpha^{\prime}}\delta_{\omega^{\prime} P_{-}} B^{\beta^{\prime}}](y) \} \{{\rm Tr}[B^{\alpha}\delta_{\omega P_{-}} B^{\beta}](0) \} |0\rangle
& \equiv &
\\
& & \hspace{-30ex} 2\pi i (g^{\alpha \alpha^{\prime}}_{\perp}g^{\beta \beta^{\prime}}_{\perp}+g^{\alpha \beta^{\prime}}_{\perp}g^{\beta \alpha^{\prime}}_{\perp})\delta_{\omega \omega^{\prime}} \int \frac{dk^{+}}{2\pi} \delta^{(2)}(y^{\perp}) \delta(y^+) e^{-\frac{i}{2} k^{+}y^{-}} J_{\omega}(k^{+}, \mu) \,. \nonumber
\end{eqnarray}
The jet function, $J_\omega(k^+,\mu)$, is only a function of one component
of the usoft momentum, $k^+$, which follows from the collinear
Lagrangian containing only the $n\cdot \partial$
derivative~\cite{Bauer:2001yt}.  We can also define a usoft function
\begin{equation}\label{ufunction}
S(l^{+},\mu) \equiv \int \frac{dy^-}{4\pi} e^{-il^{+}y^{-}} \frac{\langle 0 | \chi^{\dagger}_{-p} \boldsymbol{\sigma}^{k} \psi_{p}(y^-) a^{\dagger}_{\psi}a_{\psi} \psi^{\dagger}_{p} \boldsymbol{\sigma}^{k} \chi_{-p}(0) | 0 \rangle }{4m_c\langle {\cal O}^\psi_1(^3S_1)\rangle}\,.
\end{equation}
Combining Eqs.~(\ref{csbot}, \ref{Teff}, \ref{jetfunction}, \ref{ufunction}), we can get the fatorization theorem:
\begin{equation}\label{factthmII}
T_{\mu\nu}=H_{\mu\nu} \int dl^+ S(l^+, \mu) J_{\omega}(l^+-\sqrt{s}(1-\hat{x})) \,,
\end{equation}
with
\begin{equation}\label{HardI}
H_{\mu\nu} \equiv  \frac{m_c}{6\pi} \langle {\cal O}^\psi_1(^3S_1)\rangle (v^\delta v^{\delta'} - g^{\delta \delta'})(g^{\alpha \alpha^{\prime}}_{\perp}g^{\beta \beta^{\prime}}_{\perp}+g^{\alpha \beta^{\prime}}_{\perp}g^{\beta \alpha^{\prime}}_{\perp}) \Gamma^{\dagger}_{\alpha^{\prime}\beta^{\prime}\mu\delta^{\prime}}\Gamma_{\alpha\beta\nu\delta}  \,.
\end{equation}
Plugging Eq.~(\ref{factthmII}) back into Eq.~(\ref{xsec}), proves the result Eq.~(\ref{factthmI}).

Changing variables from $p_{\psi}$ to $z=E_{c\bar{c}}/E^{max}_{c\bar{c}}$ with $E^{max}_{c\bar{c}} = \sqrt{s}(1+r)/2$ and integrating over $\cos \theta$, we finally get
\begin{eqnarray}\label{xsec2}
\frac{d\sigma}{dz} &=& \frac{256\pi}{81}\frac{\alpha^2 \alpha_s^2 e_c^2}{s^2m_c}\frac{(1+r)(2r+1)}{(1-r)}\langle {\cal O}^\psi_1(^3S_1)\rangle P[r,z]\int dl^+S(l^+, \mu) J_{\omega}(l^+-\sqrt{s}(1-\hat{x}))\nonumber \\
&=&\sigma_0 P[r,z]\int dl^+S(l^+, \mu) J_{\omega}(l^+-\sqrt{s}(1-\hat{x})) \,.
\end{eqnarray}
Here
\begin{equation}
\label{sig0}
\sigma_0 = \frac{256\pi}{81}\frac{\alpha^2\alpha_s^2e_c^2}{s^2m_c}\frac{(1+r)(1+2r)}{1-r}\langle {\cal O}^\psi_1(^3S_1)\rangle
\end{equation}
is the differential cross section at the end point predicted by NRQCD and $P[r,z] = \sqrt{(1+r)^2 z^2 - 4 r}/(1-r)$ is a phase space factor. Note that $P[r,1] = 1$.

\begin{figure}[t]
\centerline{ \includegraphics[width=2.5in]{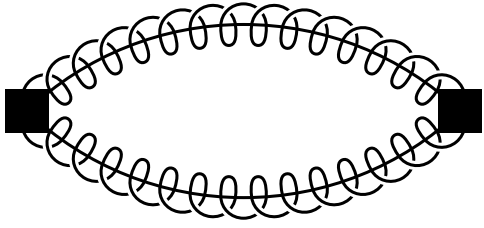}}
\caption{\it Feynman diagram for the leading order jet function.
\label{vacloop}}
\end{figure}

To leading order the jet function can be calculated easily.
 The Feynman diagram for the vacuum matrix element is shown in Fig.~\ref{vacloop}. By evaluating the one loop integral, we get
\begin{equation}\label{jetlead}
{\rm Im}J_{\omega}(k^+) = \frac{1}{2}\int^1_{-1} d \xi \delta_{\omega,\sqrt{s}(1-r)\xi} \,.
\end{equation}
Substituting it into differential cross section in Eq.~(\ref{xsec2}) and summing over $\omega$ gives,
\begin{equation}\label{xseclead}
\frac{d\sigma}{dz} = \sigma_0 P[r,z]\int dl^+ S(l^+, \mu) \Theta(l^+-\sqrt{s}(1-\hat{x})) \,.
\end{equation}
The color-singlet usoft function just shifts the endpoint from the partonic to the physical hadronic endpoint~\cite{Beneke:1997qw}.  To show this, we first note that the usoft function can formally be written as
\begin{equation}\label{soft}
S(\ell^+,\mu) = \frac{ \langle 0 | \chi^{\dagger}_{-p} \boldsymbol{\sigma}^{k} \psi_{p}
\delta(i n\cdot \partial - \ell^+) a^{\dagger}_{\Psi}a_{\Psi}
\psi^{\dagger}_{p}\boldsymbol{\sigma}^{k} \chi_{-p} | 0 \rangle }{4m_c\langle {\cal O}^\psi_1(^3S_1)\rangle}\,.
\end{equation}
Then by integrating over $\ell^+$ in Eq.~(\ref{xseclead}) gives
\begin{equation}\label{nopenouch}
\frac{d\sigma}{dz} = \sigma_0 P[r,z]
\frac{\langle 0 | \chi^{\dagger}_{-p} \boldsymbol{\sigma}^{k} \psi_{p}
\Theta[i n\cdot \partial+\sqrt{s}(1-\hat{x})] a^{\dagger}_{\Psi}a_{\Psi}
\psi^{\dagger}_{p}\boldsymbol{\sigma}^{k} \chi_{-p} | 0 \rangle }{4m_c\langle {\cal O}^\psi_1(^3S_1)\rangle} \,.
\end{equation}

Finally, writing $x$ in terms of $z$
\begin{eqnarray}
z &=& \frac{s x + M_\psi^2/x}{s+M_\psi^2} \approx 1 -\frac{1-r}{1+r}(1-x),\\
x &\approx& 1-\frac{1+r}{1-r}(1-z)\,,
\end{eqnarray}
and using the result in Ref.~\cite{Rothstein:1997ac} we get
\begin{equation}\label{lastresult}
\frac{d\sigma}{dz} =\Theta(1-z) \sigma_0 P[r,z] \,.
\end{equation}
Notice that as $z \rightarrow 1$, this coincides with the lowest order NRQCD result in the same limit.

\section{Resumming Sudakov Logarithms}
One of the main strengths of using an effective field theory is the ability to sum logarithms  using the renormalization group equations (RGEs).  Large logarithms of the ratio of well-separated scales arise naturally in perturbation theory, which can cause a breakdown of the perturbative expansion.  By matching onto an effective theory, the large scale is removed to be replaced by a running scale $\mu$.  After matching at the high scale, the operators are run to the low scale using the RGEs.  This sums all large logarithms into an overall factor, and any logarithms that arise in the perturbative expansion of the effective theory are of order one.

For $e^+e^-\to\jpsi+X$, there are logarithms of $\log(1-z)$ that appear in the perturbation series. Near the endpoint, $z\to1$, these become large, and need to be summed, which the RGEs will do for us.  For the color-singlet $^3S_1$ contribution, unlike the color-octet process~\cite{Fleming:2003gt}, these endpoint logarithms are single, not double, logarithms.  A similar situation occurs for radiative $\Upsilon$ decay~\cite{Fleming:2002sr}.  Double logarithms occur when there is an overlap of soft and collinear logs.  For the color-singlet case, the soft logarithms do not occur.  This can be seen by the fact that the usoft Wilson lines canceled out of the color-singlet matrix element.  Physically, the long-wavelength gluons do not couple to the tightly bound color-singlet $c\bar c$.

We have matched in the previous section onto the SCET color-singlet operator, which intergrates out the  large scale $\mu_H$, replacing it with a running scale $\mu$. We now run the color-singlet operator from the hard scale to the collinear scale, which sums all logarithms of $1-z$.
To run the color-singlet operator given in Eq.~(\ref{3s1op1}), we calculate the counterterm for the operator, determine the anomalous dimension, and then use this in the RGEs. Luckily, the calculation of the anomalous dimension has already been done in Ref.~\cite{Fleming:2002sr}, and we can lift the results from that paper.  The result for the resummed, differential cross-section is
\begin{equation}\label{dodzresum}
\frac{d\sigma_{\rm resum}}{dz} = \sigma_0 P[r,z] \Theta(1-z) \int^1_0 d\eta \Big[\frac{\alpha_s(\mu_c)}{\alpha_s(\mu_H)} \Big]^{2\gamma(\eta)} \,,
\end{equation}
where $\gamma$ is defined as
\begin{equation}
\gamma \equiv \frac{2}{\beta_0} \Bigg[ C_A \big[ \frac{11}{6} + (\eta^2+(1-\eta)^2) \big( \frac{1}{1-\eta}\ln \eta + \frac{1}{\eta} \ln (1-\eta) \big) \big] - \frac{n_f}{3} \Bigg] \,.
\end{equation}
To sum the large logarithms, we use the same hard scale as in Ref.~\cite{Fleming:2003gt}, $\mu_H=(s/M)(1-r)$ and the collinear scale $\mu_c \approx \sqrt{1-z} \mu_H$ in the above expression.\footnote{The hard scale $\mu_H$ that we use is different than the choice of Ref.~\cite{Lin:2004eu}.  However, numerically they are almost the same, and will not have a large effect on the results.}

\begin{figure}[t]
\centerline{ \includegraphics[width=4.5in]{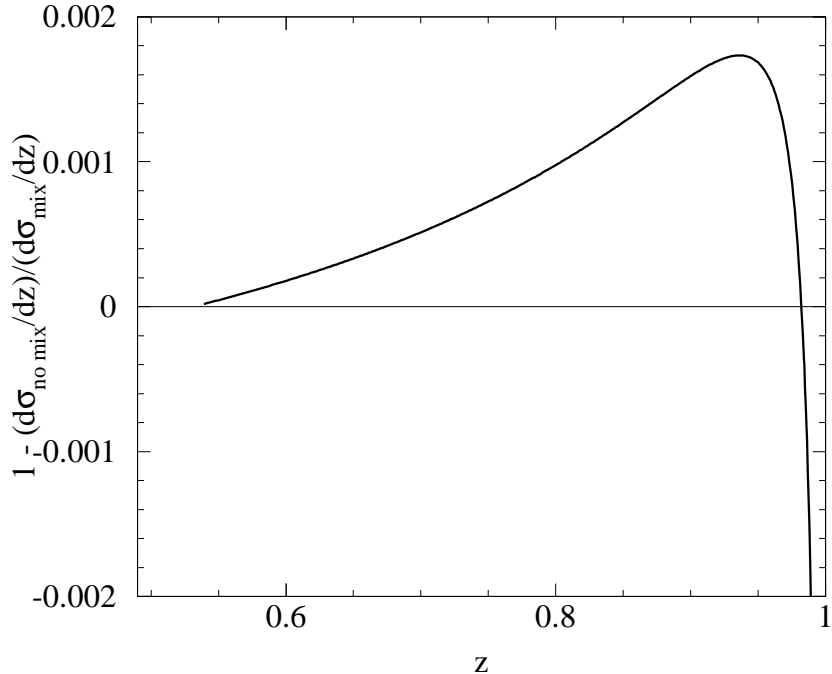}}
\caption{\it The difference between mixing and non-mixing $d\sigma_{\rm{resum}}/dz $, normalized to the mixing result, calculated at the scale $\mu_c = \sqrt{1-z} \mu_H$.
\label{mixingz}}
\end{figure}

To be completely consistent, we should include the mixing of the $gg$ jet with the $\bar q q$ jet.  Since the match onto the $\bar q q$ operator begins at a higher order than the $gg$ operator, except for very close to $z=1$ the mixing term is small~\cite{Fleming:2004rk}. The calculation of the mixing in SCET was first done in Ref.~\cite{Fleming:2004rk}, and we just quote the results here. Once we included the mixing effect, the resummed differential cross section becomes

\begin{eqnarray}\label{mixing}
\frac{1}{\sigma_0}\frac{d\sigma_{\rm{resum}}}{dz}=\frac{8}{9}P[r,z]\Theta(1-z) \sum_{\rm{n}{}\rm{odd}} & \Bigg[ &  \frac{1}{f^{(n)}_{5/2}} \Big( \gamma^{(n)}_+ r(\mu_c)^{2\lambda^{(n)}_+ / \beta_0} -\gamma^{(n)}_- r(\mu_c)^{2\lambda^{(n)}_- / \beta_0}\Big)^2  \nonumber \\
& + & \frac{3f^{(n)}_{3/2}}{8[f^{(n)}_{5/2}]^2}\frac{\gamma_{gq}^{(n)2}}{\Delta^2} \Big( r(\mu_c)^{2\lambda^{(n)}_+ / \beta_0} - r(\mu_c)^{2\lambda^{(n)}_- / \beta_0} \Big)^2 \Bigg] \,,
\end{eqnarray}
where $r(\mu)$ is defined as
\begin{equation}
r(\mu) = \frac{\alpha_s(\mu)}{\alpha_s(\mu_{\rm{H}})}
\end{equation}
and
\begin{eqnarray}
& f^{(n)}_{5/2} & = \frac{n(n+1)(n+2)(n+3)}{9(n+3/2)} \\
& f^{(n)}_{3/2} &= \frac{(n+1)(n+2)}{n+3/2} \,.
\end{eqnarray}
We also defined $\lambda^{(n)}_{\pm}$ and $\gamma^{(n)}_{\pm}$as
\begin{eqnarray}
& \lambda^{(n)}_{\pm} &= \frac{1}{2} \big[ \gamma^{(n)}_{gg} + \gamma^{(n)}_{q\bar{q}} \pm \Delta \big] \\
& \gamma^{(n)}_{\pm} &= \frac{\gamma^{(n)}_{gg}-\lambda^{(n)}_{\mp}}{\Delta} \,,
\end{eqnarray}
with
\begin{eqnarray}
&\Delta &= \sqrt{(\gamma^{(n)}_{gg}-\gamma^{(n)}_{q\bar{q}})^2 + 4 \gamma^{(n)}_{gq}\gamma^{(n)}_{qg}}\\
&\gamma^{(n)}_{gg}&= C_A\left[ \frac{2}{n(n+1)}+\frac{2}{(n+2)(n+3)}-\frac{1}{6}-2\sum_{i=2}^{n+1}\frac{1}{i}\right] -\frac{1}{3}n_f \\
&\gamma^{(n)}_{gq}&= C_F\frac{1}{3}\frac{n^2+3n+4}{(n+1)(n+2)}\\
&\gamma^{(n)}_{qg}&=3n_f\frac{n^2+3n+4}{n(n+1)(n+2)(n+3)}\\
&\gamma^{(n)}_{q\bar q}&=C_F \left[ \frac{1}{(n+1)(n+2)}-\frac{1}{2}-2\sum_{i=2}^{n+1}\frac{1}{i} \right] \,.
\end{eqnarray}
In Fig.~\ref{mixingz}, we plot the difference of the mixing result, Eq.~(\ref{mixing}), and the non-mixing result, Eq.~(\ref{dodzresum}), normalized to the mixing result.  For this plot,  we chose the scale $\mu_c = \sqrt{1-z} \mu_H$. The difference between the two is a fraction of a percent, except extremely close to the endpoint, where our results no longer hold.  We can therefore use either the mixing or the non-mixing result, Eq.~(\ref{dodzresum}) or Eq.~(\ref{mixing}).  

\section{Phenomenology}
The result from the previous section, Eq.~(\ref{dodzresum}) or Eq.~(\ref{mixing}), summed up the leading logarithmic corrections which are important near the endpoint.  Away from the endpoint, the logarithms that we have summed are not important and contributions that we neglected in the endpoint become important.  We therefore would like to interpolate between the leading order color-singlet calculation away from the endpoint and the resummed result in the endpoint.  To do this, we will define the interpolated differential rate as
\begin{equation}\label{fulleq}
\frac{1}{\sigma_0}\frac{d \sigma_{\rm int}}{d z}=
\bigg( \frac{1}{\sigma_0}\frac{d \sigma_{\rm LO}^{\rm dir}}{d z} - P[r,z] \bigg)
+ \frac{1}{\sigma_0}\frac{d \sigma_{\rm resum}}{d z} \,.
\end{equation}
The term in parentheses vanishes as $z \to 1$, leaving only the resummed contribution in that region.\footnote{This choice of interpolating between the results is different than the one made in Ref.~\cite{Lin:2004eu}.  Given the fact that the function $P[r,z]$ is a phase-space factor, we believe our choice more accurately encompasses the deviation due to higher-order QCD corrections.  The choice in interpolating factor is the largest difference between our result and the result of Ref.~\cite{Lin:2004eu}.  Note that the choice made in Eq.~(\ref{fulleq}) switches from the leading-order result to the resummed result closer to the endpoint than the choice in Ref.~\cite{Lin:2004eu}.} Away from the endpoint the resummed contribution combines with the $-P[r,z]$ to give higher order in $\alpha_s(\mu_H)$ corrections.

\begin{figure}[t]
\centerline{ \includegraphics[width=4.5in]{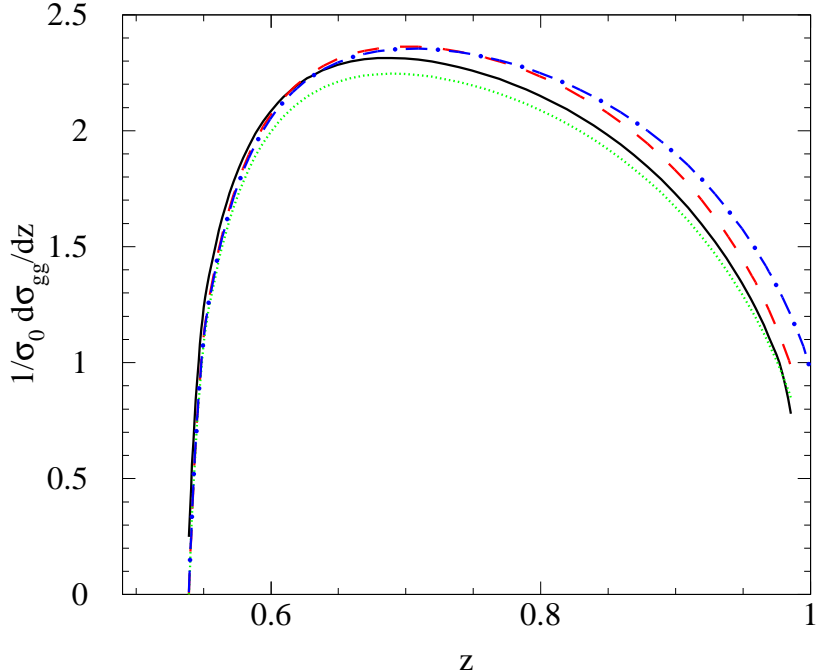}}
\caption{\it The color-singlet differential cross section . The dot-dashed curve is the leading-order NRQCD preciction.  The solid curve is the interpolated result, Eq.~(\ref{fulleq}) prediction at calculated at the scale $\mu_c = \sqrt{(1-z)} \mu_H$. The dashed curve is the interpolated result at the scale $\mu_c = 2\sqrt{(1-z)} \mu_H$, and the dotted curve is the interpolated result using the scale $\mu_c = \sqrt{(1-z)} \mu_H/2$.
\label{dsec}}
\end{figure}

For our figures, we will use $m_c = 1.4$ GeV and $\sqrt{s} = 10.58$ GeV.  In Fig.~\ref{dsec}, we compare the resummed, interpolated result, Eq.~(\ref{fulleq}), to the leading-order $e^+e^-\to J/\psi gg$ color-singlet result \cite{Cho:1996cg}.  We also show the scale dependence of the interpolated result.  The dot-dashed curve corresponds to the leading-order color-singlet result.  All curves are normalized to $\sigma_0$ given in Eq.~(\ref{sig0}).  The solid curve is the interpolated result, plotted at a scale $\mu_c = \sqrt{(1-z)} \mu_H$.  The dashed curve is the interpolated result at a scale $\mu_c = 2\sqrt{(1-z)} \mu_H$, while the dotted curve uses the scale $\mu_c = \sqrt{(1-z)} \mu_H/2$.  As can be seen, there is not a large scale dependence.  

\begin{figure}[t]
\centerline{ \includegraphics[width=4.5in]{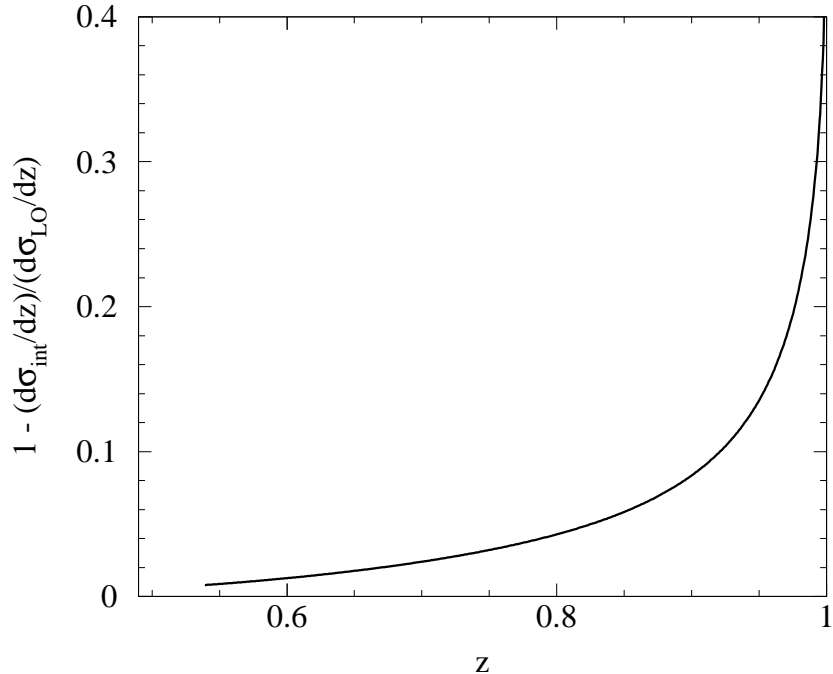}}
\caption{\it The difference of the leading-order NRQCD $e^+e^-\to\jpsi gg$ differential cross section and the interpolated result, Eq.~(\ref{fulleq}), normalized to the leading-order result.  
The interpolated result was calculated at the scale $\mu_c = \sqrt{(1-z)} \mu_H$.
\label{intratio}}
\end{figure}

As shown in Fig.~\ref{dsec}, the resummed result is smaller than the leading order result.  In order to better see the effects of the resummation, in Fig.~\ref{intratio}, we plot the difference of the leading-order, color-singlet result and the interpolated result, normalized to the leading-order result.  As can be seen, in the endpoint region there corrections become large.  However, over most of phase space, the corrections are less than 10\%.   

\begin{figure}[t]
\centerline{ \includegraphics[width=4.5in]{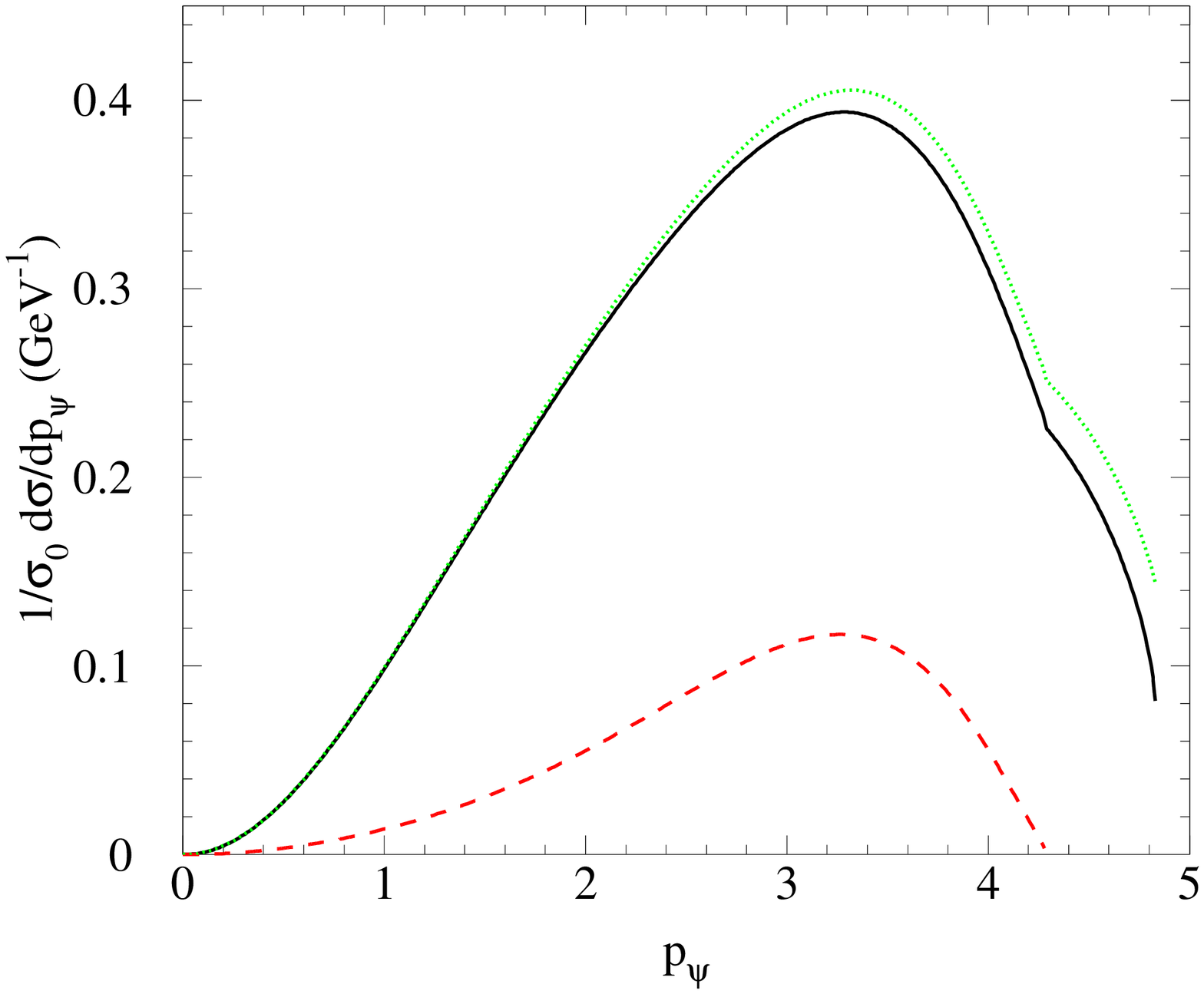}}
\caption{\it Comparison of the leading-order and resummed total color-singlet results.  The dashed curve is the NRQCD prediction for $e^+e^-\to\jpsi c\bar c$.  The dotted line is the total leading-order, color-singlet NRQCD prediction, while the solid curve is the total color-singlet prediction including the interpolated $e^+e^-\to\jpsi gg$ result.  The resummed  result was calculated at the scale $\mu_c = \sqrt{(1-z)} \mu_H$.
\label{cscomp}}
\end{figure}

The total color-singlet contribution also has the $J/\psi+c+\bar c$ final state, so we need to combine the results above with the color-singlet $e^+e^-\to J/\psi+c\bar c$ contribution~\cite{Cho:1996cg}.  In Fig.~\ref{cscomp} we compare the total leading-order, color-singlet result (dotted line) to the total, resummed color-singlet result (solid line) for $(1/\sigma_0)d\sigma/dp_\psi$.  Also shown as the dashed line is the $J/\psi+c+\bar c$ contribution. While the resummed result is slightly suppressed compared to the leading-order result, qualitatively the plots are the same.  Note that this implies that the resummation of the color-singlet contribution are not big enough to explain the anomalously large contribution to $J/\psi$ associated with extra $c\bar c$ found at the $B$ factories~\cite{Abe:2002rb,Aubert:2005tj}.

\begin{figure}[t]
\centerline{ \includegraphics[width=4.5in]{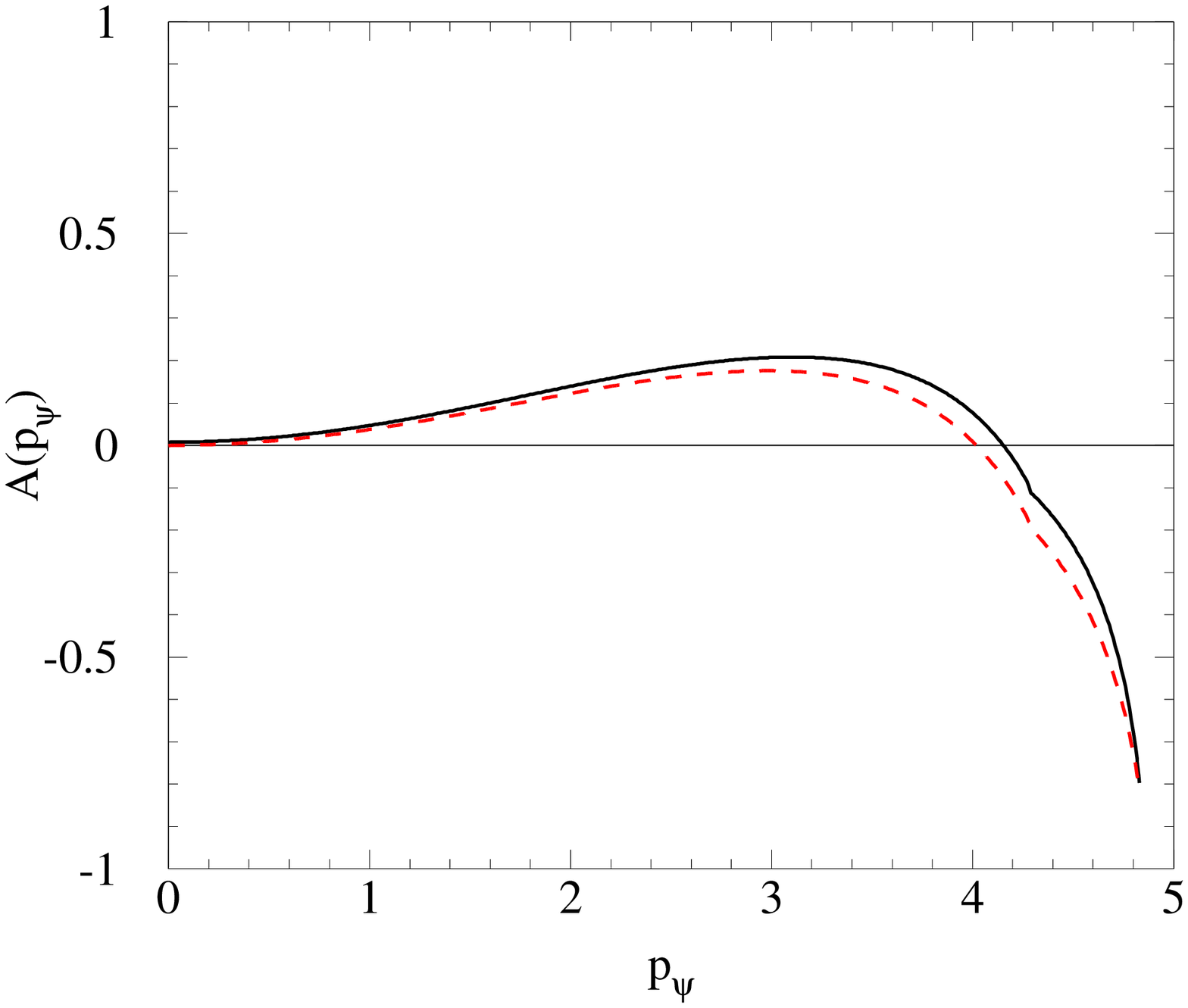}}
\caption{\it The color-singlet contribution to $A(p_{\psi})$. The solid curve is the SCET prediction, with $\mu_c = \sqrt{(1-z)} \mu_H$ and the dashed curve is the lowest-order NRQCD prediction.
\label{AP}}
\end{figure}

In Fig.~\ref{AP}, we plot the color-singlet prediction for $A(p_{\psi})$.  The dashed curve is the leading order, color-singlet result, and the solid curve is the interpolated result, including the $J/\psi +c+\bar c$ contribution.  Since the resummation is independent of the angle, both curves drop to the same value at the endpoint,
\begin{equation}
A(p_\psi^{\rm max}) = \frac{s - m_\psi^2}{s+m_\psi^2}.
\end{equation}
Away from the endpoint, the resummed color-singlet rate is slightly larger than the leading-order rate.  However, to explain the data, we still need to include the color-octet contribution.

To make a prediction for the differential cross section, we need to combine the color-singlet results discussed in this paper with the resummed color-octet results from Ref.~\cite{Fleming:2003gt} .  Given the size of the corrections found in this paper, the results are qualitatively the same as those presented in Ref.~\cite{Fleming:2003gt}.

\section{Conclusion}

In this paper, we study the color-singlet contribution to $\jpsi$ production in $e^+ e^-$ collision near the kinematic end point by using a combination of SCET and  NRQCD. The calculation consists of matching onto a color-singlet operator in SCET which integrates out the hard scale. By decoupling the usoft modes from the collinear modes using a field redefinition, we are able to show a factorization theorem for the differential cross section.  The differential rate can be factorized into a hard piece, a collinear jet function, and an usoft function.   As pointed out by Ref.~\cite{Rothstein:1997ac} the usoft function in this case can be calculated, resulting in just a shift from the partonic to the physical endpoint.

By running the resulting rate from the hard scale to the collinear scale, we sum the logarithms of the ratio of the hard and colliear scales, which correspond to large Sudakov logarithms of $1-z$.   Finally, we combine the SCET calculation with the leading order, color-singlet NRQCD result to make a prediction for the color-singlet contribution to the differential cross section over the entire allowed kinematic range.  If we combine the results for the color-singlet calculation given in this paper the resummed results for the color-octet calculation given in Ref.~\cite{Fleming:2003gt}, we now have a consistent prediction over the entire kinematic range for the $e^+e^-\to J/\psi+X$ differential cross section.  

To be consistent the resummation of the color-singlet presented here must be included.  However, except for right near the endpoint the size of the corrections are small.  The color-octet contributions, as can be seen from Ref.~\cite{Fleming:2003gt}, are necessary to get a reasonable fit to the data and are larger than the color-singlet contribution over all of phase space.  Therefore, while the quantitative picture changes slightly, the qualitative picture is the same with or without running as what was presented in Ref.~\cite{Fleming:2003gt}.  In particular, we still do not have an explanation for the unexpectedly large number of $J/\psi$ being produced with extra charm.  The solution to this puzzle will have to come from another source.

\acknowledgments
We would like to thank Sean Fleming and Thomas Mehen for helpful discussions.  A.K.L.~and X.L.~were supported in part by the National Science Foundation under Grant No.~PHY-0546143.  Adam Leibovich is also supported in part by the Research Corporation.


\end{document}